\documentclass[sigconf]{acmart}

\usepackage{graphicx}
\usepackage{stfloats}

\copyrightyear{2026}
\acmYear{2026}
\setcopyright{cc}
\setcctype{by-nc-nd}
\acmConference[DIS Companion '26]{Designing Interactive Systems Conference}{June 13--17, 2026}{Singapore, Singapore}
\acmBooktitle{Designing Interactive Systems Conference (DIS Companion '26), June 13--17, 2026, Singapore, Singapore}
\acmDOI{10.1145/3802974.3808681}
\acmISBN{979-8-4007-2632-3/2026/06}

\begin{document}

\title[Fostering Emotional Perspective-Taking]{Fostering Emotional Perspective-Taking: An Exploration of Affective Face-Tracking Interactions in the VR Narrative Rekindle}

\author{Hector Fan}
\email{fan.huaf@northeastern.edu}
\orcid{0009-0009-4303-3818}
\affiliation{%
  \institution{Northeastern University}
  \city{Boston}
  \state{MA}
  \country{USA}
}

\author{Casper Hartveld}
\email{c.harteveld@northeastern.edu}
\affiliation{%
  \institution{Northeastern University}
  \city{Boston}
  \state{MA}
  \country{USA}
}

\author{Mark Sivak}
\email{m.sivak@northeastern.edu}
\affiliation{%
  \institution{Northeastern University}
  \city{Boston}
  \state{MA}
  \country{USA}
}

\begin{abstract}
Interest in leveraging emotions in Interactive Digital Narrative (IDN) has been growing, and Virtual Reality (VR) offers rich access to real-time biometric data such as facial expressions; yet this capability remains underexplored in novel IDN design. Existing approaches typically treat emotion input superficially, such as adjusting system difficulty or aesthetics, but rarely influence how players experience the narrative itself. Prior work also lacks a focus on a specific authored narrative. We propose an experimental affective interaction model that uses a VR headset's built-in face-tracking capability to recognize player emotional states, fostering ``emotional perspective-taking'' between the player and their embodied story character, thereby deepening the player's emotional connection to the character and their narrative engagement with the VR experience.\enlargethispage{12pt}
\end{abstract}




\begin{teaserfigure}
\centering
\includegraphics[width=\textwidth]{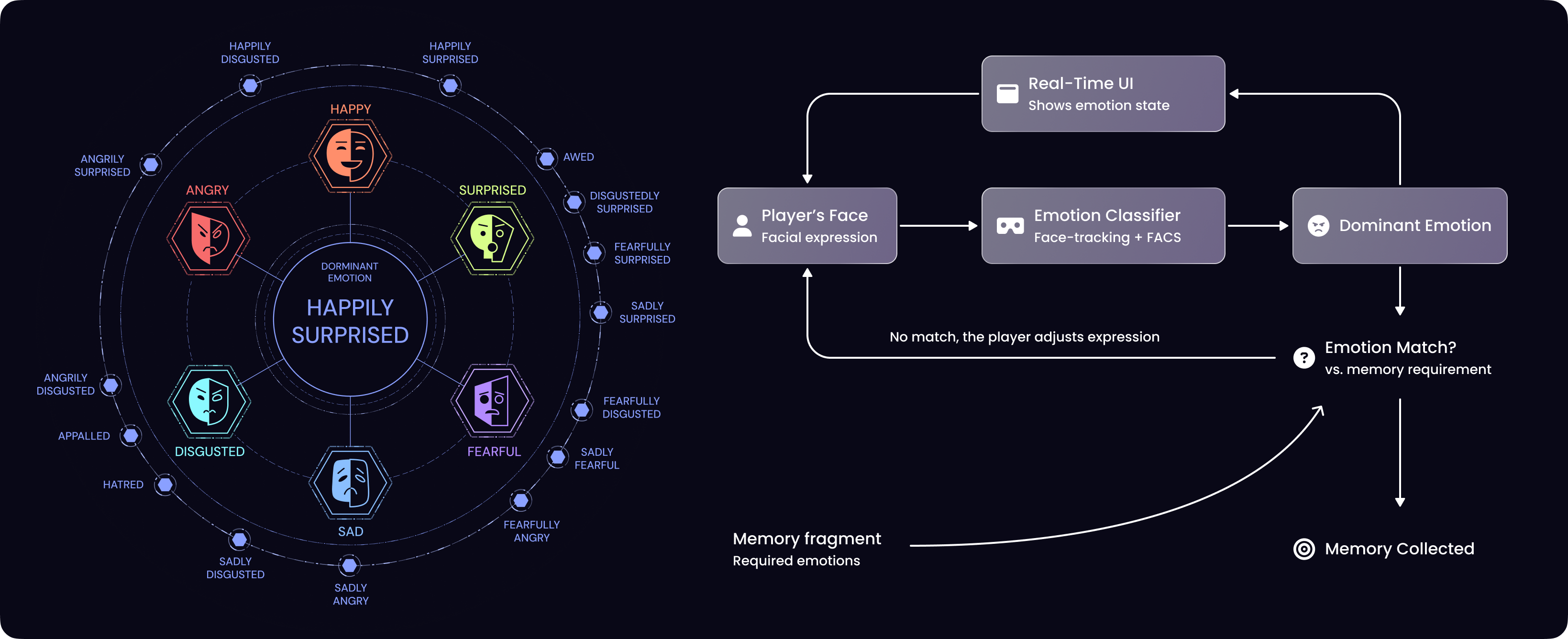}
\caption{Affective emotional perspective-taking interaction in Rekindle. Left: a within-experience UI reflects the player's real-time facial emotions. Right: the interaction flow, in which the player's facial expression is classified through a face-tracking and FACS-based method; if it matches the memory fragment's required emotion, the player can proceed to collect the memory; if not, the player adjusts their expression and retries.}
\label{fig:teaser}
\end{teaserfigure}

\maketitle

\section{Introduction}

\enlargethispage{12pt}

\looseness-1Murray~\cite{Murray_HamleHolod_2017} argues that \textit{dramatic agency}, not mere interactivity, should be the goal of Interactive Digital Narrative (IDN). Unlike common understandings of agency that equate it with the availability of options to a player, dramatic agency emerges when interactions are driven by anticipation of story events, which allow players to act in ways that feel meaningfully connected to the narrative. Recent scholarship on agency has shifted focus from providing players with affordances for choice to supporting \textit{perceived agency}. Kway and Mitchell~\cite{Kway_PerceAgenc_2018} argue that perceived agency arises from the ``player's ability and willingness to engage in meaningful expression of the playable character's personality within constraints,'' and that such agency is further supported when the system recognizes this meaningful expression through appropriate feedback. Crucially, even in games with more choices, players may treat interactions as ``just clicking through the game'' rather than engaging with the character.

\looseness-1\textit{Rekindle}~\cite{Fan_Rekindle_2024} is a first-person Virtual Reality (VR) IDN set in a dystopian future where a regime enforces heteronormativity through memory manipulation. It was designed to exemplify dramatic agency. The player steps into the shoes of the protagonist and embarks on a journey to retrieve lost memories and reassert his identity. Dramatic agency is exemplified through a \textit{memory retrieval mechanic}: players actively search for memory fragments scattered throughout the virtual environment, a process that grants them agency for exploration while coherently embodying the protagonist's journey to recover lost memories and reassert his identity. These memory fragments are unchangeable story events disseminated throughout the virtual environment and presented through animated 3D characters, accompanied by dialogue, at different moments.

\looseness-1A prior empirical study of Rekindle revealed that some players rush through memory fragments without letting animations and dialogue finish, treating the retrieval process as a game quest rather than engaging with the narrative meaning embedded within them, which likely reduces their narrative comprehension and emotional engagement with the story. This exhibits precisely the lack of engagement Kway and Mitchell~\cite{Kway_PerceAgenc_2018} describe, suggesting that narrative comprehension failures arise not from insufficient choices but from insufficient emotional engagement. We hypothesize that interactions requiring players to emotionally engage with the story character to retrieve lost memories---creating an alignment between ludological goal and emotional participation---will increase perceived agency and deepen narrative engagement and comprehension.

\looseness-1These findings present a design opportunity: could affective interactions foster deeper emotional engagement by actively influencing players' emotional states during the memory retrieval process? Specifically, we hypothesize that requiring players to engage in ``emotional perspective-taking'' might allow them to interact with the narrative more intentionally, thus deepening their emotional connection with the story. Building on Rekindle's existing memory retrieval mechanic, we implemented experimental affective interactions using face-tracking that determines player emotional state in real time using the Facial Action Coding System (FACS)~\cite{Ekman_FaciaActio_1978}. To collect a specific memory fragment, players must express the corresponding emotion facially, correctly enacting the protagonist's emotions at specific events. This paper presents the design of this affective interaction model.

\begin{figure*}[htbp]
\centering
\includegraphics[width=\textwidth]{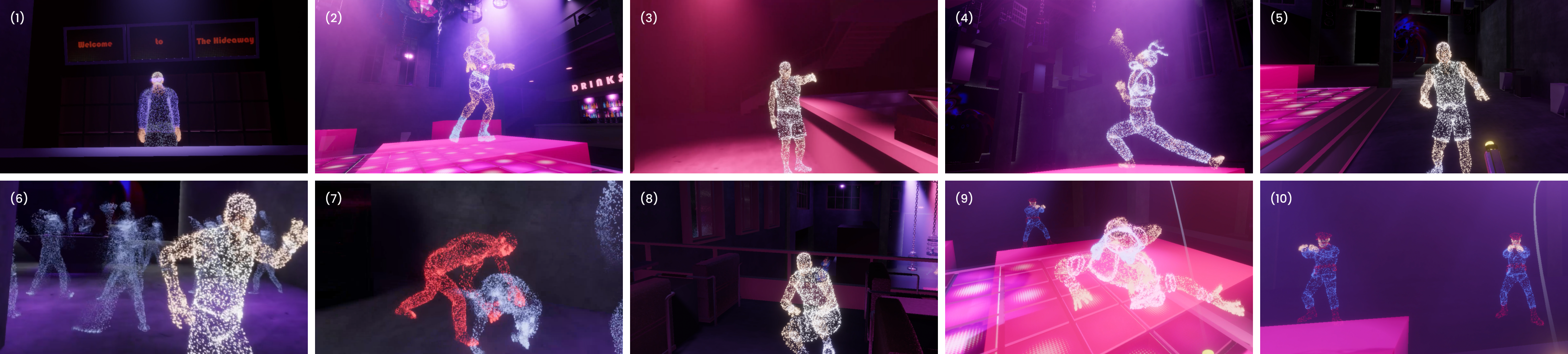}
\caption{The ten memory fragments scattered throughout Rekindle's virtual environment, presented as glowing particles of 3D characters. Each fragment is tagged with a required emotion that the player must express through facial expression to collect it.}
\label{fig:memory_fragments}
\end{figure*}

\section{Related Works}
\enlargethispage{12pt}

\subsection{Emotion Recognition in Interactive Experiences}

\looseness-1Research on emotion recognition in interactive systems spans several distinct applications. Marsella et al.~\cite{Marsella_InterPedag_2003a, Marsella_ExpreBehav_2004, Marsella_InterPedag_2003} established the importance of virtual humans displaying nonverbal cues to convey internal cognitive and emotional states, demonstrating that readable emotional expression fosters player comprehension of character motivations. This paradigm treats emotion as output: the system expresses; the player interprets.

\looseness-1Early explorations of affective interaction in IDN attempted to invert this relationship. The CALLAS project allowed players to interpret an expressive character's emotion, then use voice-based input to alter the character's actions~\cite{Charles_AffecInter_2007}. However, latency limitations in emotion recognition at that time prevented real-time emotional expression as input, constraining the design to sequential interpretation-then-action rather than continuous affective interaction.

\looseness-1More recent work has leveraged improved physiological and Facial Emotion Recognition (FER) in VR, though applications remain concentrated in training, therapy, and general VR research rather than in narrative~\cite{Halbig_SysteRevie_2021, Ortmann_FaciaEmoti_2023}. Game applications primarily use emotion to adjust difficulty or aesthetics: Houzangbe et al.~\cite{Houzangbe_FearBiofe_2018} demonstrated a horror VR game that modifies the player's field of view based on heart rate; \textit{Nevermind}~\cite{FlyingMol_Never_2015} alters game environments in response to detected stress and fear; \textit{SentimentVoice}~\cite{Ryu_SentiInteg_2025} uses face-tracking and voice analysis to shift virtual environment aesthetics to reflect participant emotions. Novel camera-based biometric inputs have also emerged. Before Your Eyes~\cite{GoodbyeWor_YourEyes_2021} uses eye-blink detection to advance gameplay, and Goodnight Universe~\cite{NiceDream_GoodnUnive_2025} extends this approach: beyond using blinks to interact with game objects, it integrates creative interactions such as closing the eyes to use powers like mind-reading and telekinesis, or allowing players to express happiness or sadness through facial expressions to react to specific narrative moments.

\looseness-1However, these applications rarely focus on a specific authored narrative---the focus of Rekindle. Emotion serves as a modifier of mechanical parameters, such as difficulty or aesthetics, rather than as a component that fosters comprehension of a specific narrative itself. Systematic reviews by Halbig and Latoschik~\cite{Halbig_SysteRevie_2021} and Ortmann et al.~\cite{Ortmann_FaciaEmoti_2023} confirm this gap: while biometric research in VR is extensive, applications focused on authored narrative remain scarce. This presents an opportunity for affective interaction designs that tie emotional input directly to narrative meaning rather than to peripheral system adjustments.

\subsection{Emotional Perspective-Taking Through Facial Expression}

\looseness-1Our approach inverts the paradigm established by Marsella et al.'s~\cite{Marsella_InterPedag_2003a} virtual human work. Rather than the system expressing emotions for the player to interpret, we require the player to express specific emotions to interact, hypothesizing that this active emotional display deepens engagement with the narrative.

\looseness-1This design draws on appraisal theory. According to the OCC model~\cite{Ortony_CogniStruc_2022}, emotions arise from appraisals of events, agents, and objects relative to one's goals and concerns. In our system, memory fragments function as emotionally tagged events. Players naturally experience emotions when encountering story events in VR, but these spontaneous responses may not align with the emotions the story character would feel. By requiring specific emotional expressions, we ask players to actively consider how the character they embody might feel at each story moment. The interaction thus becomes a mechanic for \textit{emotional perspective-taking}: players must step into the character's emotional experience, not merely observe it.

\looseness-1This approach occupies a middle ground between explicit and implicit interaction paradigms. Gilroy et al.~\cite{Gilroy_ExploPassi_2012} critiqued the dominance of explicit inputs in interactive narrative, arguing that physiological and emotional signals offer richer engagement than button presses. Knoller~\cite{Knoller_AgencArt_2010} theorized that unintentional inputs create meaning retrospectively through ``implicit causation.'' Our design incorporates both dimensions: when a player's spontaneous emotional response aligns with the intended story emotion, the system recognizes this alignment and allows natural progression, creating the implicit causation Knoller describes.

\section{Application Design}

\enlargethispage{12pt}

\subsection{Facial Emotion Recognition}

\looseness-1While biometric methods such as electromyography (EMG) and electroencephalography (EEG) can measure emotional states, they require additional sensors attached to the player. Modern VR headsets have begun incorporating more comprehensive face-tracking capabilities that track not only the lower face region but also the upper face region (eyes, eyebrows) previously occluded in image-based FER research.

\looseness-1We utilize the Meta Quest Pro's FACS-based blendshapes~\cite{Meta_FaceTrack_2025}, which represent 70 facial muscle movements used to animate expressions. Rather than relying on image-based classification, we map these blendshapes to standard Action Units (AUs) and categorize emotions using Du et al.'s~\cite{Du_CompoFacia_2014} compound emotion model. This model provides 21 emotion categories (excluding neutral), including 6 basic emotions and 15 compound emotions such as \textit{happily surprised}, \textit{sadly angry}, and \textit{fearfully disgusted} (Figure~\ref{fig:teaser}, left). This expanded taxonomy enables more nuanced recognition than basic emotion classification alone. The system returns a value between 0 and 1 indicating the strength of each emotion category.

\subsection{Face-tracking Calibration}

\begin{figure}[htbp]
\centerline{\includegraphics[width=1.0\linewidth]{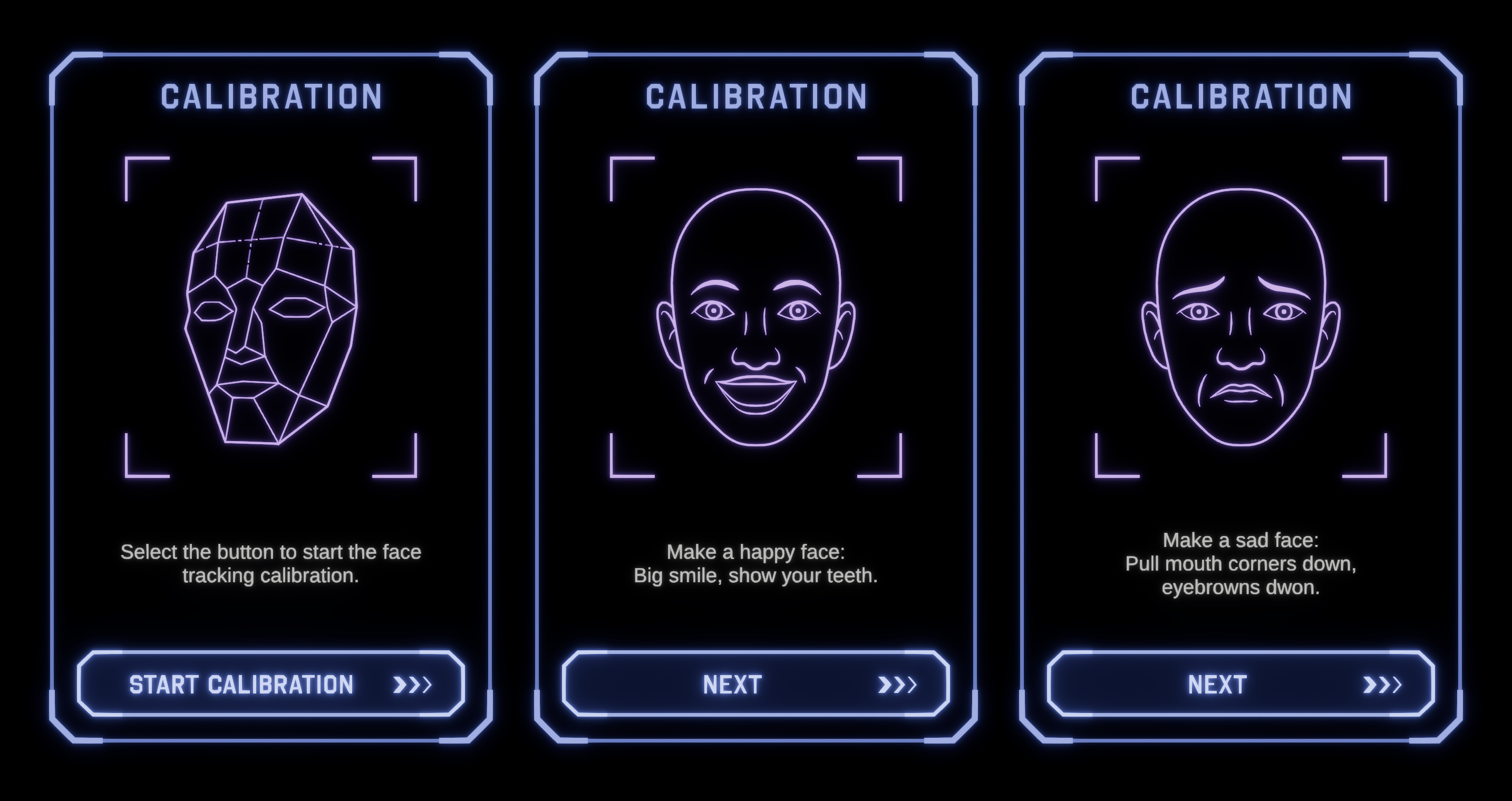}}
\caption{The face-tracking calibration initiates at the beginning of each session, in which players are guided to perform seven exaggerated facial expressions, for instance, the happy face shown in the middle and the sad face shown on the right.}
\label{fig:Calibration}
\end{figure}

\looseness-1Individual players vary in the intensity of facial expressions they naturally produce, which could result in friction if a player's subtle expressions fail to register. To address this, we designed a calibration process at the start of each session. Players are asked to perform several exaggerated facial expressions, and the calibration algorithm maps the maximum value achieved for each blendshape to 1.0 (Figure~\ref{fig:Calibration}). This normalization adjusts sensitivity to each player's expressive range, improving emotion categorization accuracy across different users.

\subsection{Interaction Design}

\looseness-1The original memory retrieval mechanic allows players to trigger a memory event when they are within a proximity trigger at a set distance, and the Memory Illuminator allows the player to collect each memory once it is triggered. In comparison, the new interaction model asks players to conduct ``emotional perspective-taking,'' meaning they have to express their emotions through facial expressions that match the predetermined emotional profiles corresponding to each memory fragment to successfully collect the memory (Figure~\ref{fig:teaser}, right).

\begin{figure*}[t]
    \centering
    \includegraphics[width=1\textwidth]{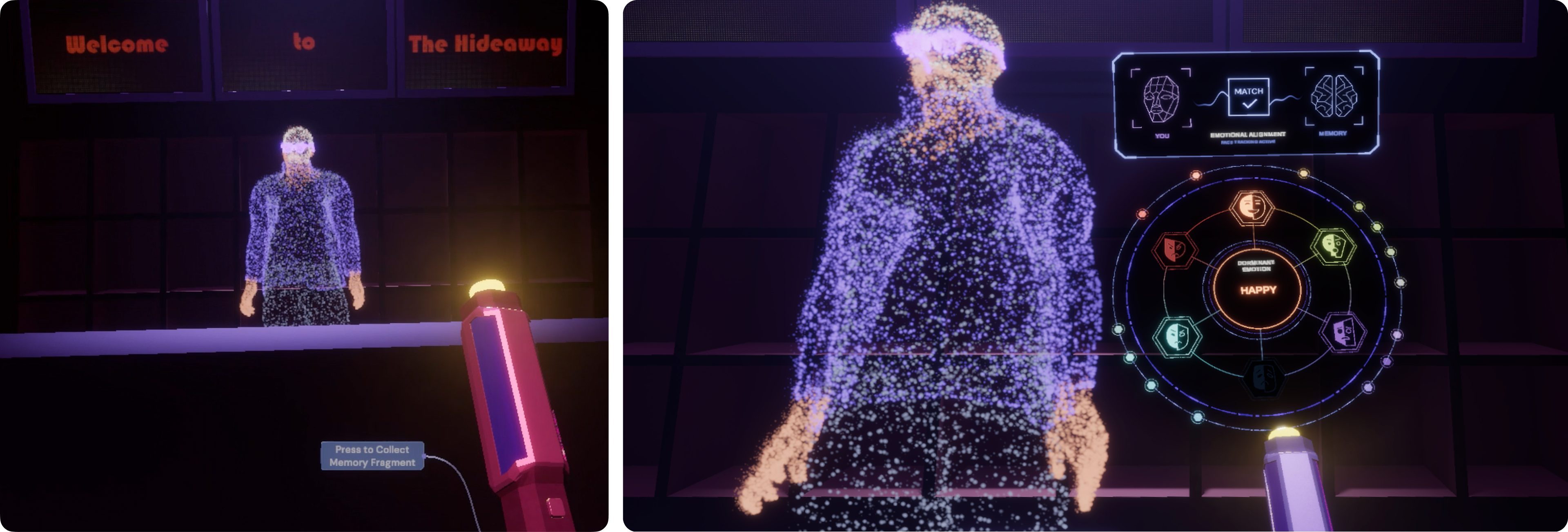}
    \caption{Comparison of the original memory retrieval mechanic (left), and the affective interaction model (right), which displays the player's real-time emotional state and requires a matching emotion to collect the fragment.}
    \label{fig:interaction_comparison}
\end{figure*}

\looseness-1In this model, each memory fragment is treated as a story event that may trigger a natural emotional response. We assigned required emotions to all ten memory fragments (Figure~\ref{fig:memory_fragments}) based on narrative context. For example, the memory \textit{Owen Greeting} requires \textit{happy}, \textit{surprised}, or \textit{happily surprised}; the memory \textit{Police Raiding} requires \textit{angry}, \textit{fearfully angry}, or \textit{fearful}. When the player's dominant emotion matches one of the required emotions, they can collect that memory fragment.

\looseness-1A user interface (UI), shown on the right side of Figure~\ref{fig:interaction_comparison}, displays the player's real-time emotional state, with the current dominant emotion highlighted in the center. A surrounding radial UI displays all 21 emotion categories, 6 basic emotions on the inner circle and 15 compound emotions on the outer, using icon brightness to convey intensity. Above this display, an emotional alignment indicator shows whether the player's dominant emotion matches the emotion(s) required by the memory fragment. Crucially, the required emotion(s) are intentionally hidden from the UI, encouraging the player to interpret the narrative meaning of the memory in order to infer them, rather than simply performing toward a stated goal. To subtly guide the player, a waveform fluctuates in intensity to reflect how closely the player's dominant emotion aligns with the memory's required emotion(s). When a match is achieved, a match icon appears, signaling that the player can now collect the memory.

\looseness-1This design serves two purposes. First, it prevents players from rushing through memory collection as though it were simple button-clicking; the required emotions function as ``keys'' that gate progression. Second, it encourages emotional perspective-taking: players must consider what the character they embody might have felt at each narrative moment, rather than approaching the experience from their own perspective.

\subsection{Preliminary Research Design}

\begin{figure*}[b]
    \centering
    \includegraphics[width=1\textwidth]{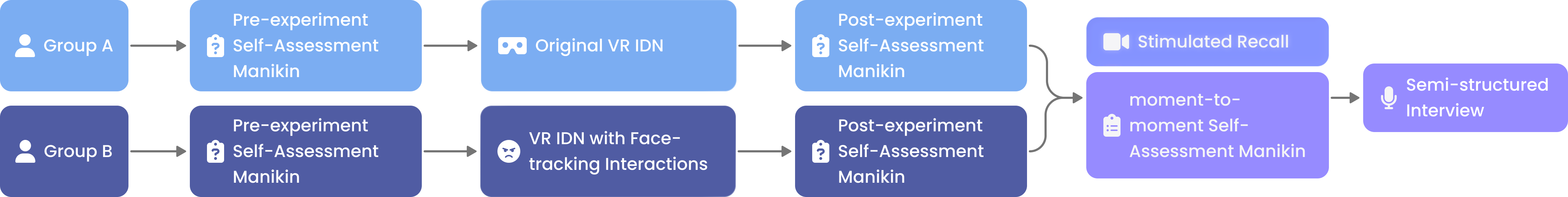}
    \caption{Proposed between-subjects study procedure comparing the original Rekindle with the affective version.}
    \label{fig:study_procedure}
\end{figure*}

\looseness-1To study how our experimental affective interactions influence players' engagement with the VR experience, we propose a preliminary research design that compares the original Rekindle with the affective version (Figure~\ref{fig:study_procedure}).

\looseness-1Participants first complete a Self-Assessment Manikin (SAM) survey instrument to evaluate their pre-experiment emotional state (including valence and arousal). They are then randomly assigned to one of two groups to play either the original VR experience or the affective version. During the VR experience, participants' facial emotions from both groups will be recorded throughout. After the VR experience, participants complete the SAM survey again to evaluate their post-experiment emotional state. Participants then view recorded walkthrough videos of their own sessions and evaluate their emotional state at each of the 10 memory events (Figure~\ref{fig:memory_fragments}). Finally, one-on-one semi-structured interviews will be conducted to evaluate participants' narrative comprehension.

\looseness-1This design allows for comparing players' pre- and post-experiment emotional states across the two conditions, examining the affective interaction's effects at specific story events within the VR experience to determine whether the intended "emotional perspective-taking" allows players to align with or deviate from the intended story event emotions. The comparison of the two groups' narrative comprehension interviews will provide insights into whether the affective interactions foster better comprehension.
\section{Conclusion and Future Works}

This paper extends the original VR IDN Rekindle with face-tracking-based affective interactions, inviting players to conduct emotional perspective-taking with their embodied story character through facial expression. We position this work as an early exploration of a broader affective interaction design principle for VR IDN with authored story: the player's affective enactment serves as both interaction input and part of meaning-making itself. When a player's spontaneous emotional response aligns with the intended emotional state in the story, this enactment proceeds implicitly; when it diverges, the mechanic invites the player to intentionally express the intended emotion. We argue that affective interaction represents a critical opportunity for VR IDN: by integrating emotional signals as interaction input beyond controller-based input alone, it makes players' emotional participation part of the narrative meaning-making process and creates meaningful agency that enriches their engagement with the narrative system.

To study how such affective interactions influence players' narrative experience, we plan to conduct a between-subjects study that compares the original Rekindle with the affective interaction version, examining differences in emotional engagement and narrative comprehension. Additionally, while the current design of the affective interaction invites players to conduct emotional perspective-taking, it does not actively cue players, especially when players' spontaneous emotional response diverges from the intended story emotion. A promising direction for future work is to integrate a large language model (LLM)-driven story character that processes the narrative state and the player's response in real time, offering subtle in-world dialogue cues to guide players' emotional states toward the intended story emotions without breaking narrative immersion.

\begin{acks}
This demo was developed on the foundation of the original Rekindle project. The first author experimented with incorporating face-tracking-based affective interactions into the original memory retrieval mechanism in the Affective Computing course at Northeastern University. We sincerely thank Professor Stacy Marsella for his invaluable feedback on developing the affective interactions and designing the initial research study.
\end{acks}

\bibliographystyle{ACM-Reference-Format}
\bibliography{References}

\end{document}